\documentclass[11pt,twoside]{article}
\usepackage{./asp2014}

\aspSuppressVolSlug
\resetcounters

\def\water{H$_2$O}
\def\kms{km\,s$^{-1}$}
\def\kmsmpc{km\,s$^{-1}$\,Mpc$^{-1}$}

\def\H0{{\it H$_0$}}
\def\LCDM{$\Lambda$CDM}

\begin{document}

\title{H$_2$O Megamaser Cosmology with the ngVLA}
\author{Jim Braatz,$^1$ Dom Pesce,$^{1,2}$ Jim Condon,$^1$ and Mark Reid$^3$
\affil{$^1$National Radio Astronomy Observatory, Charlottesville, VA 22903}
\affil{$^2$University of Virginia, Charlottesville, VA 22903}
\affil{$^3$Harvard-Smithsonian Center for Astrophysics, Cambridge, MA 02138}}

\paperauthor{Jim Braatz}{jbraatz@nrao.edu}{}{NRAO}{}{Charlottesville}{VA}{22903}{USA}
\paperauthor{Dom Pesce}{dpesce@virginia.edu}{}{UVa}{Astronomy}{Charlottesville}{VA}{22903}{USA}
\paperauthor{Jim Condon}{jcondon@nrao.edu}{}{NRAO}{}{Charlottesville}{VA}{22903}{USA}
\paperauthor{Mark Reid}{reid@cfa.harvard.edu}{}{SAO}{}{Cambridge}{MA}{02138}{USA}


\section{Description of the problem}
The expansion rate of the universe at z $\simeq$ 0, known as the Hubble
Constant (\H0), is the most fundamental parameter in 
observational cosmology.  Historically, measurements of \H0\ have been sought
to reveal basic properties of the universe, including its age and geometry.
Measuring \H0\ remains a primary focus of observational cosmology today.  The 
goal of new studies with the ngVLA will be to determine \H0\ to $\sim$1\% by
measuring geometric distances to \water\ megamasers directly in the Hubble
Flow.  A measurement at this precision is required to improve our
understanding of dark energy and to test the validity of the 
standard \LCDM\ model of cosmology.

\section{Scientific importance}

The exquisite observations of the Cosmic Microwave Background (CMB) by the 
WMAP and Planck satellites set a powerful framework for
precision cosmology.  These observations determine 
the angular-size distance to the surface of last scattering at z $\simeq$ 1100
and constrain the geometry of the universe to be very nearly flat.
The CMB, however, does not uniquely determine all fundamental cosmological 
parameters on its own.  Observations at z $\simeq$ 0, when dark energy is 
dominant, provide complementary data that constrain critical 
parameters, including the dark energy equation of state, 
neutrino mass, and the number of families of relativistic particles. 
CMB observations can {\it predict} basic cosmological parameters, including 
\H0, but only in the context of a specific cosmological model. 
Comparing CMB predictions to astrophysical measurements of \H0\ therefore 
provides a powerful test of cosmological models.

In the context of the standard model of cosmology, i.e. a geometrically flat 
\LCDM\ universe, Planck measurements predict \H0\ = 67.8 $\pm$ 0.9 \kmsmpc\ 
\citep{planck16}. 
Measurements from Baryon Acoustic Oscillations combined with SN~Ia determine 
H$_0$ = 67.3 $\pm$ 1.0 km s$^{-1}$ Mpc$^{-1}$ 
\citep{alam17},
in line with the Planck prediction.  Measurements anchored at low redshift 
and based on standard candle techniques, however, are in tension with the 
BAO measurements and the Planck predictions: 
H$_0$ =73.24 $\pm$ 1.74 km s$^{-1}$ Mpc$^{-1}$ 
\citep{riess16} and H$_0$~=~74.3~$\pm$~2.6~km~s$^{-1}$~Mpc$^{-1}$ 
\citep{freedman12}.
Determinations based on observations of gravitationally lensed quasars,
meanwhile, give H$_0$~=~71.9~$^{+2.4}_{-3.0}$~km~s$^{-1}$~Mpc$^{-1}$ 
\citep{bonvin17}.

The measurements of \H0\ based on low-z observations and the predictions based
on \LCDM\ disagree at high significance, and understanding this discrepancy 
is of fundamental importance.
Either the underlying observations suffer from unrecognized uncertainties, 
against all scrutiny, or adjustments are needed to the \LCDM\ 
standard model of cosmology.  A problem of such fundamental importance 
warrants new, independent, and percent-level measurements of \H0.  The 
long-term goal for 
the observational cosmology community is to reach percent-level precision 
in \H0\ with agreement across multiple, independent observational methods.

Measuring \H0\ with \water\ megamasers is a powerful complement to standard
candle methods because it is a one-step
measurement, independent of distance ladders and calibrations, and it is a
fundamentally geometric method that measures \H0\ directly at z~$\simeq$~0.

\section{Astronomical impact}

Understanding the cosmological model and the nature of dark energy is perhaps
the most important problem in fundamental physics.  It is
unclear whether dark energy is an intrinsic property of the universe
(the cosmological constant), or related to a time-variant field with a
yet-undiscovered particle.   Modifications to General Relativity,
the introduction of new relativistic particles, and identification of new 
modes of interaction between matter and radiation are all also viable options 
to resolve the outstanding tension between measurements and the standard model.

Refinements to the cosmological model would impact our understanding of all 
aspects of the universe on its largest scales, including remnant background 
radiation, the formation of structure, and galaxy and cluster evolution.

\section{Anticipated results}

An ngVLA measurement of \H0\ can build on methods and results from the
Megamaser Cosmology Project (MCP), a multi-year, international project 
to measure \H0\ using the megamaser technique \citep{reid13}.  The project
measures distances to suitable megamaser systems by fitting a warped disk
model to observations of 22 GHz water vapor megamasers in the nuclear accretion 
disk of the host AGNs.  Figure 1 shows an example of observations of an
edge-on accretion disk megamaser system in the nucleus of the Seyfert 2 galaxy NGC 5765b.  
Systematic errors in the disk modeling are not likely to be 
correlated among different galaxies, so the final measurement of \H0\ is 
determined as a weighted mean of measurements to individual galaxies.  
With present-day
instrumentation, the MCP has so far measured H$_0$ = 69.3 $\pm$ 4.2 \kmsmpc\ 
\citep{braatz18}.  When complete, the project should achieve a $\sim$4\% 
total uncertainty.

With the ngVLA, the megamaser-based measurement of \H0\ will aim to 
reach $\sim$1\% total uncertainty.
If the ngVLA measurement of \H0\ aligns with the standard-candle measurements,
which are themselves improving, the evidence would become convincingly against 
the standard \LCDM\ model.  If it aligns with the Planck prediction, 
additional scrutiny would be warranted to search for systematics among the
different astrophysical measurement methods.

\section{Limitations of current astronomical instrumentation}

Megamaser systems used to measure \H0\ are faint, and the megamaser method is
limited by the 22 GHz sensitivity of the telescopes available for their study.
The MCP uses the most sensitive suite of telescopes working today at 22 GHz,
including the GBT for surveys and spectral monitoring observations, and the 
High Sensitivity Array (the VLBA, GBT, VLA, and 100-m Effelsberg telescope) 
to map maser disk systems.
The final measurement by the MCP will be based upon distances to nine
maser disk systems bright enough to measure with these existing facilities.  
Targeted surveys of over 3000 galaxies were necessary to identify those nine.

\section{Connection to unique ngVLA capabilities}

Only the ngVLA, with its sensitivity approaching an order of magnitude increase
over existing facilities, will make a 1\% measurement of \H0\ plausible using
the megamaser method.   The measurement, however, will require that the ngVLA
meet certain design features to enable precision astrometry of faint masers.
Importantly, the ngVLA must have a compact core of antennas containing a 
substantial 
fraction of the total collecting area.  This core should be concentrated within
$\sim$5 km to enable efficient phasing at 22~GHz.  

The method also requires that the ngVLA include, or have coordinated access
to, a number of VLBI stations on intercontinental baselines out to $\sim$5000 
km, ideally with substantial collecting area of their own.  
Long baselines in both the E-W and \mbox{N-S} directions are required.  
Contiguous 
frequency coverage from 18-22 GHz within a single receiver band is also 
important to make surveys efficient for detection of high velocity
maser emission.  Furthermore, flexible subarray capabilities will be 
beneficial so that, for example, outer antennas can be utilized for other 
science while the phased core is operating in a VLBI mode.

\section{Experimental layout}

Measuring \H0\ with the megamaser technique requires three types of 
observations. First is a survey to identify the rare, edge-on 
disk megamasers suitable for distance measurements. Second is sensitive 
spectral monitoring of those disk megamasers to measure secular drifts in 
maser lines, indicative of the centripetal accelerations of maser clouds as 
they orbit the central black hole.  
And third is sensitive VLBI observations to map the maser features and 
determine the rotation structure and angular size of the disk.  
For each megamaser disk being measured, the spectral monitoring would span 
1-2 years and include observations on a roughly monthly cadence.  Since each
VLBI map includes the necessary spectral information needed to track
line-of-sight accelerations, a strategy to map and monitor the maser disks
simultaneously is feasible with the ngVLA.  The positions, velocities, and 
accelerations of the maser components are then used to constrain a model of 
a warped disk and determine the distance to the host galaxy.

The precision with which the distance to a megamaser disk can be 
measured depends on a number of factors, including the richness of its
maser spectrum, the layout of observable masers within the disk, and the 
signal-to-noise of the measurements.  For megamasers in the Hubble flow, 
higher signal-to-noise observations would equate to more precise distances 
in all cases.  So, the first stage 
of the experimental layout would be to measure distances to all {\it known}
megamasers suitable for such measurements.
The MCP surveys and others have discovered $\sim$20 megamaser disk systems
whose spectral profile indicates an edge-on disk suitable for a distance
measurement.  Although the MCP will already have measured distances to nine
of those megamasers, those nine would have to be reobserved with the ngVLA
to improve their individual measurement uncertainties.

The overall measurement of \H0\ based on these $\sim$20 megamasers is not 
likely to reach the 1\% project goal, so new megamaser disks must be
discovered.  The ngVLA project could aim for 
10\% distances to $\sim$100 megamasers, or 7\% distances to $\sim$50.  
In practice, the precision for each system is not well known until the disk is 
observed and modeled.  The prototypical megamaser in NGC\,4258, at a distance
of 7.54 $\pm$ 0.17 $\pm$ 0.10 Mpc, demonstrates that 
systematics in an individual system can be $<$ 3\% with 
sufficient sensitivity and angular resolution \citep{riess16}.

Ultimately, to reach 1\% the second stage of the experiment requires a survey.
With a nearly order of magnitude advantage in sensitivity, the ngVLA would be 
able to discover $\sim$30 times more megamasers than the GBT.  It would also
extend the practical distance to which megamasers can be measured to a few 
hundred Mpc, alleviating potential concerns about local variations in the 
Hubble parameter.

\section{Complementarity}

Dark energy is the most important unsolved problem in modern physics, and 
an accurate local measurement of \H0\ is the most effective complement to 
CMB data for constraining the equation of state of dark energy.  A number of
future facilities, for example WFIRST, will continue to focus on
investigations of dark energy and cosmology.

A precise and independent measurement of the Hubble constant also elevates 
the effectiveness of future CMB observations, such as the ``Stage-4'' 
ground-based experiment, CMB-S4.  Besides dark energy, a percent-level prior on
\H0\ also provides the best complement to CMB-S4 experiments for
constraining the neutrino mass \citep{manzotti16}.

The importance of the Hubble 
constant demands agreement and verification using several measurement 
techniques. The long-term goal for observational cosmology is to achieve
percent-level measurements that are consistent across independent methods, 
to minimize the impact of systematic uncertainties.

\begin{figure}
\begin{center}
\includegraphics[width=5in]{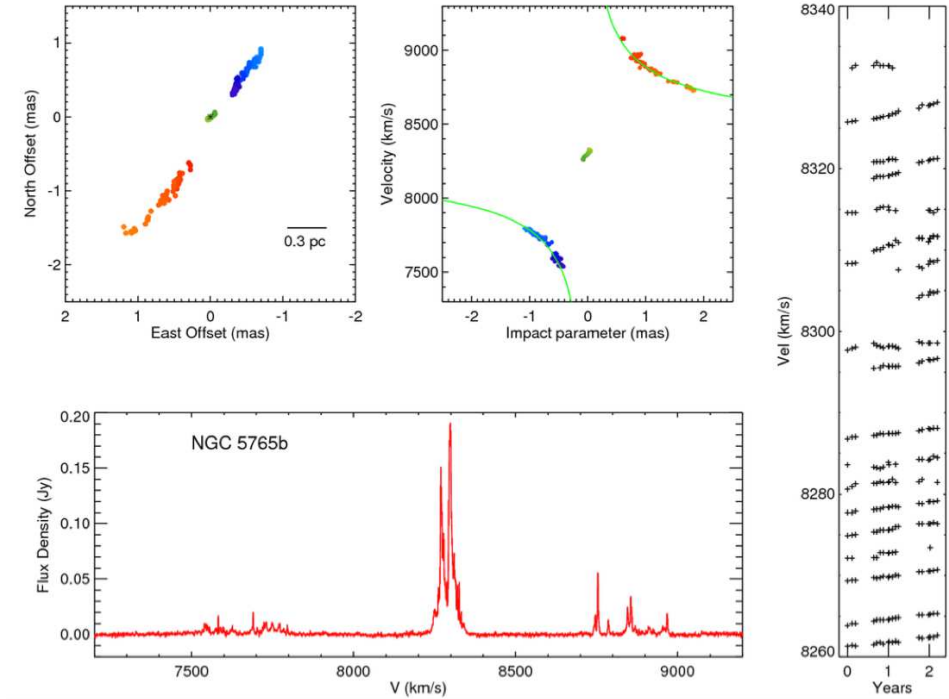}
\caption{The H$_2$O Megamaser in NGC 5765b (also see \cite{gao16}).  
The maser observations in this galaxy determine its distance at
126.3 $\pm$ 11.6 Mpc.  The top left panel shows the VLBI
maser map, with colors of the maser spots representing the line-of-sight
velocities.   The top center panel shows a position-velocity (P-V) diagram.  
The impact parameter represented on the x-axis is the angular distance measured
along the length of the edge-on disk.  The solid green lines 
on the P-V diagram represent a Keplerian fit to the rotation curve.  The 
bottom panel shows a representative GBT spectrum.  Masers in edge-on AGN 
accretion disks have a characteristic profile with three groups of maser 
features, evident in this spectrum.  The features centered near 8300 \kms\ originate from the
front side of the disk while those centered near 7650 \kms\ and 8850 \kms\ originate from the 
approaching and receding sides of the edge-on disk, respectively.  The right panel shows results of GBT
spectral monitoring.  Each symbol on the plot marks the velocity of a maser 
peak in the systemic part of the spectrum.  The maser velocities increase 
with time and represent the centripetal 
acceleration as maser clouds orbit the central supermassive black hole.}
\label{fig1}
\end{center}
\end{figure}




\end{document}